\documentclass[a4paper]{report}
\usepackage[utf8]{inputenc}
\usepackage[T1]{fontenc}
\usepackage{RJournal}
\usepackage{amsmath,amssymb,array}
\usepackage{booktabs}
\usepackage{algorithm,algcompatible,amsmath}
\algnewcommand\INPUT{\item[\textbf{Input:}]}%
\algnewcommand\OUTPUT{\item[\textbf{Output:}]}%

\begin{document}

\sectionhead{Contributed research article}
\volume{XX}
\volnumber{YY}
\year{20ZZ}
\month{AAAA}

\begin{article}
\title{\pkg{RMM}: An  R  Package for Customer Choice-Based Revenue Management Models for Sales Transaction Data}
\author{by  Chul Kim, Sanghoon Cho, Jongho Im}

\maketitle

\abstract{ 
We develop an R package \CRANpkg{RMM} to implement a Conditional Logit (CL) model using the Robust Demand Estimation (RDE) method introduced in \citet{cho2020robust}, a customer choice-based \textbf{R}evenue \textbf{M}anagement \textbf{M}odel. In business, it is important to understand customers' choice behavior and preferences when the product prices change over time and across various customers. However, it is difficult to estimate actual demand because of unobservable no-purchase customers (i.e., truncated demand issue). The CL model fitted using the RDE method,  enables a more general utility model with frequent product price changes. It does not require the aggregation of sales data into time windows to capture each customer's choice behavior. This study uses real hotel transaction data to introduce the R package \CRANpkg{RMM} to provide functions that enable users to fit the CL model using the RDE method along with estimates of choice probabilities,  size of no-purchase customers, and their standard errors.} 

\section{1 \quad Introduction}
\label{sec:intro} 
There is an extensive history of research on techniques used to estimate product or service demand across a wide area of industries. However, in practice, it is very challenging to estimate demand because of unobservable no-purchase customers (i.e., truncated demand issue). Although customers may visit a physical store or online website, they may not purchase products from the offered portfolios because of the following reasons:(1) their willingness-to-pay is lower than the offered prices; (2) the case of unintentional product out-of-stocks  (or sell-outs); (3) the case of intentionally forcing lower-price options to be unavailable \citep{ferguson2020estimating}. Unobserved no-purchase customers provide a distorted view of demand, resulting in a lower estimated mean and variance. Basing demand estimates on censored data will result in biased forecasts for future demand,  which will lead to a profit spiral down.

To resolve this issue, the Revenue Management Model (RMM) has been developed. Specifically, Conditional Logit (CL) modeling has been receiving attention in academia and practice. \citet{talluri2004revenue} developed the expectation-maximization (EM) method to estimate parameters associated with purchase probability and arrival rates, based solely on observed sales data. Some other methods have been proposed, including a variant of the EM method \citep{vulcano2012estimating}, two-step estimation method \citep{newman2014estimation} and maximize-minorize optimization method \citep{abdallah2021demand}. However, these methods require a data aggregation process for product attributes including the prices and choice sets by each time window. They also require the use of a cell  mean utility model, assuming that the product attributes remain constant across customers and time periods. Namely, it is difficult to fit choice models by capturing the varying product attributes and available choice sets over time. This leads to a loss of information and potentially biased estimates. As an alternative, we use the Robust Demand Estimation (RDE) method proposed by \citet{cho2020robust}, which is a customer choice-based RMM and enables a more general utility model with frequent price changes in the products. Additionally, it does not require aggregating sales data into time windows to capture each customer's choice behavior. The ability to manage frequent price changes is vital for demand estimation, especially in the areas of airlines,  hotels, and e-commerce, where dynamic pricing is prevalent.  

To apply the RDE method to censored data, \citet{cho2020robust} employs a CL  model to fit the censored transaction data. Although the parameters of the CL model for the available products can be consistently estimated using conventional maximum likelihood estimation, the no-purchase utility cannot be estimated without further information. \citet{cho2020robust} considered the following two  additional types of information to identify the model parameters: 1) additional assumptions on customers' utility function, and 2) external information about a firm's market share. Afterwards, \citet{cho2020robust} developed robust estimation algorithm to address the inaccuracies in the information type and let the data decide the most appropriate approach.

Various CL modeling approaches are already accessible in R, such as, for instance, \CRANpkg{mlogit} \citep{mlogit}, \CRANpkg{Rchoice} \citep{sarrias2016discrete}, and \CRANpkg{mixl} \citep{molloy2021mixl}. However, there is no suitable R package to implement the handling of censored sales data by employing the CL model. To the best of our knowledge, other publicly available software, including SAS and Python, do not provide this function either.   

The source codes of the \CRANpkg{RMM} package are accessible in \cite{RMMdocu}. It is available after installation in R 4.1.0 or later versions. There are three main functions in \CRANpkg{RMM}. First, the function \code{rmm\_reshape()} should be used by all users to fit the CL  model, by changing the data to a wide format and defining information such as the response and alternative specific variables from the given data. The function \code{rmm()} fits a CL model using the RDE method. This function  estimates the parameters, their standard error and size of the no-purchase. Finally, the function \code{predict()} produces a prediction value, which is the choice probability for each alternative in the fitted model. The customer's decision can be confirmed by this prediction value.

\section{2 \quad Conditional Logit Model using the Robust Demand Estimation Method}
The CL model using the robust demand estimation proposed by \citet{cho2020robust} uses the concept of random utility maximization. It captures each customer's choice behavior where customer $i$ is assumed to choose product $j$ from the choice set $S_{i}$ having the maximum utility. The choice set $S_i$ can include up to $J$ products in total. Consider the mean utility of customer $i$ for product $j$, 
\begin{equation}\label{fullmodel}
\nu_{ij}=\alpha_{j}+ \boldsymbol{\beta}\mathbf{x}_{ij},\; i=1, \ldots, n,\; j\in S_{i} \notag 
\end{equation}
where $\alpha_{j}$ is a fixed utility associated with the product $j$, $\mathbf{x}_{ij}$ is a vector attributes  of product $j$ exposed to customer $i$, $\boldsymbol{\beta}$ is a vector of regression slope coefficients corresponding to the product attributes $\mathbf{x}_{ij}$. Utilities can be vary across customers of the same product.

The CL model is a common RMM derived by assuming random utility $U_{ij}$,
\begin{equation}\label{RU}
U_{ij} = \nu_{ij} + \epsilon_{ij} \notag 
\end{equation}
where $\nu_{ij}$ are mean utilities (systematic component) and $\epsilon_{ij}$ are random errors (i.e., an unobserved component) which follow \textit{i.i.d.} the Gumbel distribution. Under this assumption, the choice probability for product $j$ for a customer $i$ and the no-purchase probability are respectively given as:
\begin{equation}
p_{ij} = \frac{\exp(\nu_{ij})}{1+\sum_{j \in S_i}\exp(\nu_{ij})},%
\mbox{
and }p_{i0}=1-\sum_{j \in S_i}p_{ij},  \label{eqn:choiceprob}
\end{equation}
Using the choice probabilities in (\ref{eqn:choiceprob}), the complete log-likelihood is given by 
\begin{equation}\label{eq:fulllike}
l_{comp}(\boldsymbol{\alpha}, \boldsymbol{\beta})=\sum_{i=1 }^n \left\{(1-\delta_i)\log{p_{i0}}+\sum_{j \in S_i}\delta_{ij}\log{p_{ij}} \right\},  
\end{equation}
where $\boldsymbol{\alpha}=(\alpha_1, \dots, \alpha_J)$, $\delta_{ij}$ represents the indicator function if customer $i$ purchases product $j$ and $\delta_i=\sum_{j \in S_i}\delta_{ij}$, for $i = 1, \ldots, n$. Because the no-purchase  records are unobservable, the complete-data likelihood function (\ref{eq:fulllike}) should be written again in a reduced form, initially discussed in \cite{mcfadden1973conditional}, as follows:
\begin{equation}\label{eq:clike}
l_{obs}(\boldsymbol{\alpha}^*,\boldsymbol{\beta})=\sum_{i=1}^n \delta_i \sum_{j \in S_i}\delta_{ij}\log \left( \frac{\exp(\alpha^*_{j}+\boldsymbol{\beta} \mathbf{x}^*_{ij})}{\sum_{j \in S_i}\exp(\alpha^*_{j}+\boldsymbol{\beta} \mathbf{x}^*_{ij})}\right),  
\end{equation}
where $\boldsymbol{\alpha}^\ast=(\alpha^\ast_1, \dots, \alpha^\ast_J)$ and $\mathbf{x}_{ij}^\ast$ are normalized parameters and covariates, respectively, defined as 
\begin{eqnarray}
\alpha^*_j&=&\alpha_j-\alpha_k, \label{cond1} \notag \\ 
\mathbf{x}^*_{ij}&=&\mathbf{x}_{ij}-\mathbf{x}_{ik},  \label{cond2} \notag
\end{eqnarray}
for a fixed baseline product $k \in S_i$. For parameter identification, we reduce the dimension of $\alpha$ from $K$ to $K-1$ possible choices. Similarly, the covariate $\mathbf{x}$, sharing the same model parameter $\boldsymbol{\beta}$, is normalized to be compatible across different sets of available products. To find the fixed baseline product $k$ with the smallest value among all $\alpha_k$ values, \citet{cho2020robust} developed a grid searching algorithm presented in \ref{algo:searching}.
\begin{algorithm}[!h]
	\caption{}\label{algo:searching}
	\begin{algorithmic}[1]
		\INPUT  $S$: a set of products.
		\OUTPUT Baseline product $k^*$
		\STATE \textbf{for} $k \in S$ \textbf{do}
		\STATE Obtain MLE $\hat{\boldsymbol{\eta}}^*_{-k}$ by maximizing (\ref{eq:clike})
		\STATE \quad \textbf{if} $\hat{\alpha}^*_j \ge 0$ for all $j\ne k$ \textbf{then}
		\STATE \quad \quad $k^*=k$
			\STATE \quad \textbf{else} $k^*=j$, where $\hat{\alpha}^*_{j}$ satisfies $\hat{\alpha}^*_{j} \le \hat{\alpha}^*_{l}$ for all $l \in S$.
		\STATE \quad \textbf{end if}
	\end{algorithmic}
\end{algorithm}

\citet{cho2020robust} defined $\it{instant~loss~rate}$, evaluated at purchase by the customer $i$, as the odds of no purchase probability given by
\begin{eqnarray}\label{def:lr}
l_i&=&\frac{p_{i0}}{1-p_{i0}} = \frac{1-p_{i}}{p_{i}}= \frac{\exp{(\nu_{i0})}}{\sum_{j \in S_i}\exp{(\nu_{ij})}}.
\end{eqnarray}
where the utility of customer $i$ from no-purchase is $\nu_{i0}$ and probability of purchasing at least one of the available products is $p_{i}=\sum_{j\in S_{i}}p_{ij}$. The instant loss rate in (\ref{def:lr}) is a relative ratio of no-purchase against purchase given choice set and product attributes, according to $S_i$.
Thus, it can be understood as the number of no-purchase customers exposed to the same choice set with customer $i$. Because customers are independent of each other but share the same choice model, the loss rate on customer $i$ is the expected number of no-purchase customers. Accordingly, the total number of no-purchase customers, indicated by $L$, in the observed time period can be defined as
\begin{equation}\label{eq:nopur}
L=\sum_{i=1}^n \delta_i l_i. \notag
\end{equation}

Considering the choice model parameters and definition of instant loss rate (\ref{def:lr}), we can rewrite the total number of no purchases as 
\begin{equation}\label{eq:nopur2}
L=\exp(\gamma)\sum_{i=1}^n \delta_i \left\{\sum_{j \in S_i}\nu_{ij}(\boldsymbol{\eta}^*_{-k})\right\}^{-1},
\end{equation}
where $\gamma=-\alpha_k$ is the model parameter corresponding to the baseline product $k$, and $\boldsymbol{\eta}^*_{-k}$ is the normalized vector of model parameters excluding $\alpha_k$. However, the parameter estimate $\gamma$ is not generally identifiable and estimable from the observed log-likelihood (\ref{eq:clike}). Thus, we cannot obtain a consistent estimator of $L$ despite the consistent maximum likelihood estimates $\hat{\boldsymbol{\eta}}^*_{-k}$. To estimate no-purchase utility $\gamma$, \citet{cho2020robust} considered additional information of market share as applied in \cite{vulcano2012estimating} and \cite{abdallah2021demand}. To identify and estimate $\gamma$, we incorporate market share information $s$ where $s \in (0,1)$. We construct an estimation function in which the ratio between the size of no-purchases and purchases is equal to the inversed odds of the market share, that is, 
\begin{equation}\label{mcond}
\frac{\text{No purchases (L)}}{\text{Purchases $(n_R)$}}=\frac{1-s}{s}.
\end{equation}
From the Equations (\ref{eq:nopur2}) and (\ref{mcond}), we derive the estimation of $\gamma$,
\begin{equation}\label{const}
U(\gamma \mid \hat{\boldsymbol{\eta}}^*_{-k})=\frac{1}{n_R}\exp(\gamma)\left[\sum_{i=1}^n \delta_i \left\{\sum_{j \in S_i}\nu_{ij}(\hat{\boldsymbol{\eta}}^*_{-k})\right\}^{-1}\right]-\frac{1-s}{s}=0. \notag
\end{equation}
The asymptotic properties of the MLE estimates $\hat{\boldsymbol{\eta}}^*_{-k}$ obtained by maximizing the observed log-likelihood (\ref{eq:clike}) with the baseline product $k$, are well-known  along with the likelihood theory. The variance of $\hat{\boldsymbol{\eta}}^*_{-k}$ can be estimated using the Hessian of the observed log-likelihood (\ref{eq:clike}). The asymptotic properties of the no-purchase parameter estimator $\hat{\gamma}$ are presented in Theorem 1 in \citet{cho2020robust}.

\section{3 \quad Implementation of the \CRANpkg{RMM} package}

In this section, we explain how to fit the CL model using the RDE method discussed in the previous section, through the data of \code{Hotel\_Long} and \code{Hotel\_Wide}, contained in the \CRANpkg{RMM} package. Usually, two formats of customers' transaction data are generated from a hotel, airline, or e-commerce field, namely, long or wide format. The long format data corresponding to \code{Hotel\_Long} records the attributes of each alternative in a single row (many rows are required to represent each customer's decision situation. The wide format data corresponding to \code{Hotel\_Wide} records each customer's decision situation in a single row (this format requires more columns than the long format). Here, we start by introducing the sources of the two data (\code{Hotel\_Long} , \code{Hotel\_Wide}).

\subsection{ \texorpdfstring{\code{Hotel\_Long} and \code{Hotel\_Wide}}{Hotel\_Long and Hotel\_Wide} data}

\code{Hotel\_Long} and \code{Hotel\_Wide} are preprocessing datasets derived from the publicly available ”Hotel 1” data introduced in \citet{bodea2009data}. In the “Hotel 1” data, customers are exposed to several types of rooms, that is, the choice sets are made available to them at the time of their visit. The room prices are recorded according to the characteristics of the customers (for example, party size or VIP status) and date. \citet{cho2020robust} used this data to study robust demand estimation, \citet{berbeglia2021comparative} and \citet{subramanian2021demand} used it to develop their revenue management models.

For illustration purposes, we performed the following preprocessing on the “Hotel 1”  data.
\begin{enumerate}
    \item Customers’ booking transactions with only one room type available in their choice set, were discarded.
    
    \item Duplicate transactions were removed.
    
    \item Choice sets with less than 33 observations representing rare cases were discarded.
    
\end{enumerate}

As a result of the above preprocessing, the data contains 1,100 transactions from 2007-02-12 to 2007-04-15(62 days). \code{Hotel\_Long} is a long format and \code{Hotel\_Wide} is a wide format of this data.

%
\code{Hotel\_Long} can be loaded as follows. \strong{Note:} 1,100 transactions are recorded in 8,318 rows because the data is in a long format. Table \ref{Table:Hotel_Long} represents the description of the 11 variables of \code{Hotel\_Long}.

\begin{example}
> library(RMM)      # Load the RMM package
> data(Hotel_Long)  # Load the long format example data
> dim(Hotel_Long)   # 8,318 observations, 11 variables
[1] 8318   11

> head(Hotel_Long)  # First 6 observations out of 8,318

  Booking_ID Purchase            Room_Type Price Party_Size Membership_Status
1         10        0          King Room 1   329          1                 1
2         10        0         Queen Room 1   329          1                 1
3         10        1          King Room 3   359          1                 1
4         10        0 2 Double Beds Room 1   359          1                 1
5         10        0         Queen Room 2   359          1                 1
6         10        0  Special Type Room 1   359          1                 1

  VIP_Membership_Status Booking_Date Check_In_Date Check_Out_Date Length_of_Stay
1                     0   2007-04-08    2007-04-09     2007-04-10              1
2                     0   2007-04-08    2007-04-09     2007-04-10              1
3                     0   2007-04-08    2007-04-09     2007-04-10              1
4                     0   2007-04-08    2007-04-09     2007-04-10              1
5                     0   2007-04-08    2007-04-09     2007-04-10              1
6                     0   2007-04-08    2007-04-09     2007-04-10              1
\end{example}

\begin{table}[!ht]
    \centering{}%
    \begin{tabular}{ll}
        \toprule
        \textbf{Name}       & \textbf{Description}   \\
        \hline
        Booking\_Id         & ID associated with a booking. \\
        \hline
        Purchase            & Indicator variable equals to one if the product identified \\
                            & is purchased by Room\_Type, zero otherwise. \\
        \hline
        Room\_Type          & Indicating a room type exposed to customers. \\ 
        \hline
        Price               & The average nightly rate paid by the customer in USD.  \\ 
        \hline
        Party\_Size          & The number of people associated with the booking.\\
        \hline
        Membership\_Status   & Status in rewards program \\
                             & (0—not a member, 1—basic, 2—elevated, 3—premium). \\
        \hline                       
        VIP\_Membership\_Status  & Membership status of a VIP rewards program member \\
                                 & (0—not a VIP, 1—basic VIP, 2—premium VIP member).\\
        \hline
        Booking\_Date      & The date the booking was created.  \\
        \hline
        Check\_In\_Date    & The date the customer checked in. \\
        \hline
        Check\_Out\_Date    & The date the customer checked out. \\ 
        \hline
        Length\_of\_Stay   & Length of stay (the number of nights). \\
        \bottomrule
    \end{tabular}
    \caption{Description of \code{Hotel\_Long} Data}\label{Table:Hotel_Long}
\end{table}

The same data as above, but converted to a wide format is \code{Hotel\_Wide} given below. 
Because one transaction is expressed in each row, 1,100 rows are recorded, 
and the number of variables is 22, which is more than that of  \code{Hotel\_Long}. Where \code{Decis\_Alts\_Code} is a numeric coded variable of \code{Room\_Type} selected by the customer, and \code{Choice\_Set} is a variable that expresses \code{Room\_Type} exposed to the customer as \code{Decis\_Alts\_Code}.

\begin{example}
> dim(Hotel_Wide)  # 1,100 observations, 22 variables
[1] 1100   22

> head(Hotel_Wide)

  Booking_ID Party_Size Membership_Status VIP_Membership_Status Booking_Date
1         22          1                 0                     0   2007-04-05
2         23          1                 0                     0   2007-04-05
3         24          1                 0                     0   2007-04-05
4         30          1                 0                     0   2007-04-05
5         32          1                 0                     0   2007-04-09
6         37          1                 3                     1   2007-04-06
  Check_In_Date Check_Out_Date Length_of_Stay            Room_Type
1    2007-04-10     2007-04-11              1          King Room 4
2    2007-04-10     2007-04-11              1          King Room 4
3    2007-04-10     2007-04-11              1  Special Type Room 1
4    2007-04-09     2007-04-11              2          King Room 4
5    2007-04-10     2007-04-11              1              Suite 1
6    2007-04-13     2007-04-14              1 2 Double Beds Room 1
  Decis_Alts_Code         Choice_Set Choice_Set_Code Price_1 Price_2 Price_3
1               5       1|5|7|8|9|10               6     399       0       0
2               5       1|5|7|8|9|10               6     399       0       0
3               8              1|5|8               7     399       0       0
4               5     1|4|5|7|8|9|10               4     359       0       0
5               9       1|4|5|8|9|10               5     399       0       0
6               1 1|2|3|4|5|7|8|9|10               2     299     279     279
  Price_4 Price_5 Price_6 Price_7 Price_8 Price_9 Price_10
1       0     439       0     399     399     499      599
2       0     439       0     399     399     499      599
3       0     439       0       0     399       0        0
4     359     399       0     359     359     459      559
5     399     439       0       0     399     499      599
6     299     339       0     299     299     379      479
\end{example}

The ``Alternative Specific Variables" (ASV) and ``Individual Specific Variables" (ISV) are present in the data in which the CL model can be fitted. The former is a variable indicating the characteristics of each product (alternatives) such as ``\code{Price}", and the latter is a variable indicating the attributes of a customer (individual) such as ``\code{Party\_Size}", ``\code{Membership\_Status}", and ``\code{VIP status}". The \CRANpkg{RMM} package can only use ASV to model customers’ choice probabilities based on the RDE method in \citet{cho2020robust}.

To apply the proposed RDE method, users must first use the \code{rmm\_reshape()} function to reshape the data into a wide format and organize the information required by the \code{rmm()} model fitting function. Once the model parameters are estimated from \code{rmm()}, the users obtain predict values of unobserved customer demand using \code{predict()}. The next subsections have detailed descriptions of each step.


\subsection{\texorpdfstring{\code{rmm\_reshape()}}{rmm\_reshape()} for preparing data}

The \code{rmm()} function for fitting the model requires an S3 class called \code{rmm\_data} as the input object, which the user can prepare using the \code{rmm\_reshape()} function. Table \ref{Table:reshape} describes arguments used in \code{rmm\_reshape()}.

\begin{example}
rmm_reshape(data, idvar, resp, alts, asv,
            alts_code = NULL, choice_set = NULL, choice_set_code = NULL,
            min_obs = 30)
\end{example}

\begin{table}[!ht]
    \centering{}%
    \begin{tabular}{llc}
        \toprule
        \textbf{Argument}    & \textbf{Explanation}  & \textbf{Default value} \\
        \hline
        \code{data}    & data frame, a long or wide format data & \\
        \hline
        \code{idvar}    & character, variable name representing  & \\
                        & each individual's id in the transaction data. & \\
        \hline
        \code{resp}       & character, variable name representing & \\
                          & result of an individual choice.  & \\
        \hline
        \code{alts}     & character vector, variable names& \\
                        & representing alternatives. & \\
        \hline
        \code{asv}      & character vector, variable names &  \\
                        & representing alternative specific variables. & \\
        \hline
        \code{alts\_code}   &  character, variable name &  \\
                            & representing numerically coded alternatives. & \code{NULL} \\
                            & \textbf{Only used when \code{data} is in wide format.} & \\
        \hline
        \code{choice\_set}  & character, variable name & \\
                            & representing a set of \code{alts\_code} &   \\
                            & exposed to individuals. & \\ 
                            & The delimiter for each alternative must be '|'.  &  \code{NULL} \\ 
                            & For example, ‘'1|2|5.'’ & \\
                            & \textbf{Used only when the \code{data} is in wide format.}& \\
        \hline
        \code{choice\_set\_code} & character, variable name & \\
                                 & representing a numerically coded choice set. & \code{NULL} \\
                                 & \textbf{Only used when \code{data} is in wide format.} & \\
        \hline
        \code{min\_obs} & numeric, specify the minimum observations  &  \\
                        & for each choice set in the transaction data. & \code{30} \\
        \bottomrule
    \end{tabular}
    \caption{Arguments to the function \code{rmm\_reshape()}}\label{Table:reshape}
\end{table}

Let us consider the case where the user has either the long or wide format data.  If the user has long format data such as \code{Hotel\_Long} and inserts it into the \code{rmm\_reshape()}, the function automatically converts the data into wide format and defines the information. Accordingly, it codes all the alternatives in the data as numbers and specifies the choice sets exposed to customers, as follows.

\begin{example}
# When data is in the long format.
rst_reshape <- rmm_reshape(data=Hotel_Long, 
                           idvar="Booking_ID", 
                           resp="Purchase", 
                           alts="Room_Type", 
                           asv="Price", 
                           min_obs=30)
\end{example}

However , if the user has wide format data such as \code{Hotel\_Wide}, three  more arguments, namely, \code{alts\_code}, \code{choice\_set}, and \code{choice\_set\_code}, must be specified when using the \code{rmm\_reshape()} function, unlike in the long format.

\begin{example}
# When data is in the wide format.
rst_reshape <- rmm_reshape(data=Hotel_Long, 
                           idvar="Booking_ID", 
                           resp="Purchase", 
                           alts="Room_Type", 
                           asv="Price", 
                           alts_code="Decis_Alts_Code",
                           choice_set="Choice_Set",
                           choice_set_code="Choice_Set_Code",
                           min_obs=30)
\end{example}

Note that, only one variable should be specified for the response variable, but ASV can specify multiple variables as a character vector in the \code{asv} argument. Here, we used ‘'\code{Purchase}'’ as the response variable and ‘'\code{Price}'’ as ASV.

The output of \code{rmm\_reshape()} is a list, which is the S3 class ‘'\code{rmm\_data}'’ required as input by \code{rmm()}.

\begin{example}
> class(rst_reshape)  # S3 class ‘'rmm_data'’
[1] "rmm_data"

> ls(rst_reshape)
[1] "Alts_Code_Desc"     "ASV"                "asv_name"          
[4] "data_wide"          "Rem_Choice_Set"     "Removed_Choice_Set"
\end{example}

The output \code{rst\_reshape\$Alts\_Code\_Desc}, represents all alternatives that exist in the transaction, by numerical coding. As shown below, our \code{Hotel\_Long} data has 10 alternatives, coded using numbers from 1 to 10.

\begin{verbatim}
> rst_reshape$Alts_Code_Desc

# A tibble: 10 x 2
   Alts_Code Room_Type           
       <int> <chr>               
 1         1 2 Double Beds Room 1
 2         2 King Room 1         
 3         3 King Room 2         
 4         4 King Room 3         
 5         5 King Room 4         
 6         6 Queen Room 1        
 7         7 Queen Room 2        
 8         8 Special Type Room 1 
 9         9 Suite 1             
10        10 Suite 2    
\end{verbatim}

The output \code{rst\_reshape\$Rem\_Choice\_Set}, shows the remaining choice sets expressed as a set of \code{Alts\_Code}. Similar to \code{Room\_Type}, each choice set is coded using numbers, starting from 1. The \code{Observation} column indicates how often each choice set was exposed in the transaction. Because we set the value of the \code{min\_obs} argument to 30, only the choice set that is exposed more than 30 times, appears here. This is why the second column is labeled as ”Remaining.”

\begin{verbatim}
> rst_reshape$Rem_Choice_Set

# A tibble: 12 x 3
   Choice_Set_Code Remaining_Choice_Set Observation
             <int> <chr>                      <int>
 1               1 1|2|3|4|5|6|7|8|9|10         150
 2               2 1|2|3|4|5|7|8|9|10            62
 3               3 1|3|4|5|7|8|9|10              75
 4               4 1|4|5|7|8|9|10               341
 5               5 1|4|5|8|9|10                  34
 6               6 1|5|7|8|9|10                  87
 7               7 1|5|8                         37
 8               8 1|5|8|9|10                    36
 9               9 2|5|8                         32
10              10 4|5|7|8|9|10                  34
11              11 4|5|8                        127
12              12 5|9|10                        85
\end{verbatim}

The removed choice sets are listed in \code{rst\_reshape\$Removed\_Choice\_Set}. These are not used in the model because they contain fewer than 30 observations. As a result of \code{rmm\_reshape()}, 34 choice sets are removed from  \code{Hotel\_Long}.

\begin{verbatim}
> rst_reshape$Removed_Choice_Set

# A tibble: 34 x 2
   Removed_Choice_Set Observation
   <chr>                    <int>
 1 1|2|3|4|5|6|7|8|10           1
 2 1|2|3|4|5|7|8|10             5
 3 1|2|3|5|6|7|8|9|10           8
 4 1|2|4|5|6|7|8|9|10          12
 5 1|2|4|5|7|8|9|10             3
 6 1|2|5|7|8|9|10               2
 7 1|3|4|5|6|7|8|9|10          26
 8 1|3|4|5|6|8|9|10             3
 9 1|3|4|5|7|8|10               8
10 1|4|5|6|7|8|9|10             9
# ... with 24 more rows
\end{verbatim}

In \code{rst\_reshape\$ASV}, user-specified ASV is stored as wide format.

\begin{verbatim}
> rst_reshape$ASV

$ASV
# A tibble: 1,100 × 10
   Price_1 Price_2 Price_3 Price_4 Price_5 Price_6 Price_7 Price_8 Price_9
     <dbl>   <dbl>   <dbl>   <dbl>   <dbl>   <dbl>   <dbl>   <dbl>   <dbl>
 1     399       0       0       0     439       0     399     399     499
 2     399       0       0       0     439       0     399     399     499
 3     399       0       0       0     439       0       0     399       0
 4     359       0       0     359     399       0     359     359     459
 5     399       0       0     399     439       0       0     399     499
 6     299     279     279     299     339       0     299     299     379
 7     399       0       0       0     439       0     399     399     499
 8     399       0       0       0     439       0       0     399       0
 9       0       0       0     319     349       0       0     319       0
10     379       0       0       0     419       0       0     379     479
   Price_10
      <dbl>
 1      599
 2      599
 3        0
 4      559
 5      599
 6      479
 7      599
 8        0
 9        0
10      579
# … with 1,090 more rows
\end{verbatim}

Data preparation for fitting the CL model is now complete. The following subsection introduces the \code{rmm()} function to fit the CL model using the RDE method developed by \citet{cho2020robust}.

\subsection{ \texorpdfstring{\code{rmm()}}{rmm()} for fitting the model}
The CL model using the RDE method discussed in the previous section is implemented through the main function \code{rmm()}. For a detailed description of arguments, see Table \ref{Table:rmm}.

\begin{example}
  rmm(rmm_data, prop=0.7)
\end{example}

\begin{table}[!ht]
    \centering{}%
    \begin{tabular}{llc}
        \toprule
        \textbf{Argument}    & \textbf{Explanation}  & \textbf{Default value} \\
        \hline
        \code{rmm\_data}    & S3 class \code{rmm\_data} which is  & \\
                            & output of \code{rmm\_reshape()} function.    & \\
        \hline
        \code{prop}        & numeric, an user-assumed market share. &  \code{0.7} \\
        \bottomrule
    \end{tabular}
    \caption{Arguments to the main function \code{rmm()}}\label{Table:rmm}
\end{table}

The \code{rmm()} function allows only the output of \code{rmm\_reshape}, \code{rmm\_data} S3 class, as an input. The \code{prop} argument refers to the market share, additional assumption,  as mentioned in Eq. (\ref{const}). Therefore, it indicates the proportion of the given transaction details to the total market transaction details. As previously discussed, in \CRANpkg{RMM} package, the customer's choice probability is modeled as a CL model of the mean utility $\nu_{ij}$, as shown in Eq. (\ref{eqn:choiceprob}).

Here, we would like to fit the conditional logit model using the robust demand estimation introduced in \citet{cho2020robust}. Examine the code below.

\begin{example}
# Fitting the conditional logit model with a market share of 0.7
rst_rmm <- rmm(rmm_data = rst_reshape, 
               prop = 0.7)
\end{example}

The result of the \code{rmm()} function is S3 object class "\code{rmm}," which uses the \code{predict.rmm()} method.

\begin{verbatim}
> class(rst_rmm)  # S3 class "rmm"
[1] "rmm"
\end{verbatim}

The first and second outputs of the \code{rmm()} function are \code{rst\_rmm\$Model}, \code{rst\_rmm\$Estimation\_Method}, respectively. They indicate that we fitted the CL model using the RDE method. In the third and fourth output, \code{rst\_rmm\$Response\_Variable}, \code{rst\_rmm\$Alternative\_Specific\_Variables} indicate the response variable and ASVs in our model. The fifth output \code{rst\_rmm\$Baseline\_Product} is the result of the Baseline Product Search Algorithm mentioned in Algorithm (\ref{algo:searching}). In our example, out of 10 alternatives, the third alternative, ‘'\code{King Room 3},'’ was searched as a baseline product. In \code{rst\_rmm\$Coefficient}, we can check the estimated values, standard errors, and P-values of each parameter in the model. Additionally, following the robust demand estimation procedure, it is possible to estimate the number of customers who have returned without purchasing products, with \code{rst\_rmm\$No\_Purchase} and \code{rst\_rmm\$Total\_Arrivals} indicating these estimates.

\begin{verbatim}
> print(rst_rmm)
$Model
[1] "Conditional Logit Model"

$Estimation_Method
[1] "Robust Demand Estimation"

$Response_Variable
[1] "Purchase"

$Alternative_Specific_Variables
[1] "Price"

$Baseline_Product
[1] 3

$Coefficients
              Estimate Std. Error z value Pr(>|z|)
gamma (-ASC3)  -3.3079     2.2766 -1.4530   0.1462
ASC1            1.3338     0.3221  4.1410   0.0000
ASC2            1.4175     0.2985  4.7487   0.0000
ASC4            2.0308     0.3148  6.4511   0.0000
ASC5            1.6915     0.4452  3.7994   0.0001
ASC6            0.4412     0.3987  1.1066   0.2685
ASC7            0.3404     0.3377  1.0080   0.3135
ASC8            0.9712     0.3158  3.0754   0.0021
ASC9            0.9756     0.7060  1.3819   0.1670
ASC10           2.4836     1.2308  2.0179   0.0436
Price          -0.0130     0.0057 -2.2807   0.0226

$Total_Arrivals_(Estimate)
[1] 1571

$Observed_Arrivals
[1] 1100

$No_Purchase_(Estimate)
[1] 471

attr(,"class")
[1] "rmm"  
\end{verbatim}

The next subsection demonstrates how to make predictions when new data are given. We use the model estimated in this subsection.

\subsection{ \texorpdfstring{\code{predict()}}{predict()} for prediction}

The \code{predict()} function allows users to obtain predictions from the estimated model. See table \ref{Table:predict} for a detailed description of the arguments.

\begin{example}
predict(object, newdata, Rem_Choice_Set, Choice_Set_Code, fixed = TRUE, ...)
\end{example}

\begin{widetable}[!ht]
    \centering{}%
    \begin{tabular}{llc}
        \toprule
        \textbf{Argument}    & \textbf{Explanation}  & \textbf{Default value} \\
        \hline
        \code{object}        & Object of class inheriting from 'rmm'.  & \\
        \hline
                                     & new data to be used for prediction. &   \\
        \code{newdata}               & Must be a data frame containing & \\
                                     & the ASVs used for model fitting. & \\
        \hline
        \code{Rem\_Choice\_Set}       & List of choice sets remaining in the data.   &  \\
        \hline
        \code{Choice\_Set\_Code}      & Specifies the choice set of new data.   & \\
        \hline
                        & If \code{fixed=TRUE}, the alternative with the highest  & \\
        \code{fixed}    & prediction probability is determined as decision. & \code{TRUE}\\
                        & Otherwise (\code{fixed=FALSE}), one of the alternatives & \\
                        & is sampled in proportion to the predictive probability.   & \\
        \hline
        \code{...} & further arguments passed to or from other methods. & \\
        \bottomrule
    \end{tabular}
    \caption{Arguments to the function \code{predict()}}\label{Table:predict}
\end{widetable}

In the \code{object} argument, we insert the output object of the rmm() function. Note that, the \code{Rem\_Choice\_Set} argument must specify the remaining choice set which is the output of the \code{rmm\_reshape()} function. As discussed earlier, this can be seen as \code{rst\_reshape\$Rem\_Choice\_Set}.

\begin{verbatim}
> Rem_Choice_Set <- rst_reshape$Rem_Choice_Set
> Rem_Choice_Set

# A tibble: 12 x 3
   Choice_Set_Code Remaining_Choice_Set Observation
             <int> <chr>                      <int>
 1               1 1|2|3|4|5|6|7|8|9|10         150
 2               2 1|2|3|4|5|7|8|9|10            62
 3               3 1|3|4|5|7|8|9|10              75
 4               4 1|4|5|7|8|9|10               341
 5               5 1|4|5|8|9|10                  34
 6               6 1|5|7|8|9|10                  87
 7               7 1|5|8                         37
 8               8 1|5|8|9|10                    36
 9               9 2|5|8                         32
10              10 4|5|7|8|9|10                  34
11              11 4|5|8                        127
12              12 5|9|10                        85
\end{verbatim}

Suppose that the new data for prediction is given as follows. According to the \newline \code{Choice\_Set\_Code}, this data assumes the situation exposed to \code{Choice\_Set\_Code} 7. Because  products numbered 1, 5, and 8 belong to \code{Choice\_Set\_Code} 7, their attributes, \code{Price} (alternative specific variable) can be seen in the new data. \code{newdata1} has five observations or five situations in which each individual is exposed to different prices.

\begin{example}
> newdata1 <- data.frame(Price_1=c(521, 321, 101, 234, 743),
+                        Price_5=c(677, 412,  98, 321, 382),
+                        Price_8=c(232, 384, 330, 590, 280))

> print(newdata1)
  Price_1 Price_5 Price_8
1     521     677     232
2     321     412     384
3     101      98     330
4     234     321     590
5     743     382     280
\end{example}

To predict the choice probability with this \code{newdata1}, the arguments of the \code{predict()} function can be specified as follows. The first output \code{rst\_pred1\$Model} indicates the type of prediction model. The second and third outputs are the prediction results, with \code{rst\_pred1\$Decision} showing the code number of the product selected from the products exposed to each individual. \code{rst\_pred1\$Probability} represents the probability of each individual choosing one of the exposed products. Because the value of the \code{fixed} argument is \code{TRUE}, the product corresponding to the highest choice probability is determined as \code{Decision}.

\begin{verbatim}
> rst_pred1 <- predict(object = rst_rmm, 
+                      newdata = newdata1,
+                      Rem_Choice_Set = Rem_Choice_Set,
+                      Choice_Set_Code = 7, 
+                      fixed = TRUE)
> print(rst_pred1)

$Model
[1] "Prdiction by Conditional Logit Model."

$Decision
[1] 8 1 5 1 8

$Probabiltiy
          Alts_1      Alts_5      Alts_8
[1,] 0.032273722 0.006073611 0.961652667
[2,] 0.573101692 0.251077881 0.175820427
[3,] 0.396453969 0.589491288 0.014054743
[4,] 0.681064726 0.314302956 0.004632318
[5,] 0.002256071 0.352244224 0.645499705
\end{verbatim}

To examine the case where \code{fixed=FALSE}, suppose that \code{newdata2} is given. These data are of three people exposed to \code{Choice\_Set\_Code} 3.

\begin{verbatim}
> newdata2 <- data.frame(Price_1=c(232, 122, 524), Price_3=c(152, 531, 221),
+                        Price_4=c(123, 743, 192), Price_5=c(139, 535, 325),
+                        Price_7=c(136, 276, 673), Price_8=c(387, 153, 454),
+                        Price_9=c(262, 163, 326), Price_10=c(421, 573, 472))

> print(newdata2)
  Price_1 Price_3 Price_4 Price_5 Price_7 Price_8 Price_9 Price_10
1     232     152     123     139     136     387     262      421
2     122     531     743     535     276     153     163      573
3     524     221     192     325     673     454     326      472
\end{verbatim}

Similar to prediction with \code{newdata1}, arguments can be set as follows. However, \code{fixed = FALSE} determines the product in proportion to each choice probability. In \code{rst\_pred2\$Probability}, it can be seen that the first customer has the highest choice probability for the fourth product (\code{Alt\_4}), which is 0.4887, but the final decision is the third product. This is the difference between \code{fixed = TRUE} and \code{fixed = FASLE }.

\begin{verbatim}
> rst_pred2 <- predict(object = rst_rmm, 
+                      newdata = newdata2,
+                      Rem_Choice_Set = Rem_Choice_Set,
+                      Choice_Set_Code = 3,
+                      fixed = FALSE)

> print(rst_pred2)

$Model
[1] "Prdiction by Conditional Logit Model."

$Decision
[1] 3 8 4

$Probabiltiy
          Alts_1       Alts_3       Alts_4      Alts_5       Alts_7      Alts_8
[1,] 0.059017764 0.0439933498 0.4887438272 0.282743035 0.0761297013 0.005475253
[2,] 0.514723134 0.0006655305 0.0003222683 0.003429207 0.0257447096 0.239374472
[3,] 0.004973822 0.0673153629 0.7478395764 0.094527281 0.0002654811 0.008598506
         Alts_9     Alts_10
[1,] 0.02792824 0.015968827
[2,] 0.21112052 0.004620159
[3,] 0.04560368 0.030876287
\end{verbatim}

\section{4 \quad Conclusion}

The \CRANpkg{RMM} is a useful package for estimating the following factors: (1) the customer's choice probability and (2) the number of no-purchase customers given censored transaction data with different choice sets and product prices exposed to each individual. Accordingly, \CRANpkg{RMM} uses a CL model with robust demand estimation procedure, introduced in \citet{cho2020robust}. To the best of our knowledge, \CRANpkg{RMM} is the only package useful for handling censored sales data by employing a CL model. \CRANpkg{RMM} package can be applied without limitation as long as the data can be fitted using the CL model even if it is not transaction data. Therefore, ASVs, which indicate the characteristics of each attribute, exist as independent variables.

The current version of the \CRANpkg{RMM} has some limitations. If multiple attributes are used to fit the model using the \code{rmm()} function, only a no-interaction model can be used. For example, if there are two ASVs as an independent variable, a linear additive model such as \code{ASV1 + ASV2}, can be used . However, if a model includes an interaction effect such as \code{ASV1 * ASV2}, it cannot be fitted. Also, a multinomial logit model, which is another popular customer's choice model, is not yet covered by \CRANpkg{RMM}. However, the second issue will be resolved in the next version of \CRANpkg{RMM}.

\section{5 \quad Funding}
This material is based upon work supported by the National Research Foundation of Korea(NRF) grant funded by the Korea government(MSIT) (No. NRF-2021R1C1C1014407).

\bibliography{RJreferences}

\address{Chul Kim\\
  Departments of Applied Statistics\\
  Yonsei University\\
  03722, Seoul\\
  South Korea}
\email{statkim7578@yonsei.ac.kr}

\address{Sanghoon Cho\\
  Neeley School of Business\\
  Texas Christian University\\
  76109, Fort Worth}
\email{sanghoon.cho@tcu.edu}

\address{Jongho Im\\
  Departments of Applied Statistics\\
  Yonsei University\\
  03722, Seoul\\
  South Korea}
\email{ijh38@yonsei.ac.kr}

\end{article}

\end{document}